\pdfoutput=1
\documentclass[preprint,preprintnumbers,amsmath,amssymb,floatfix,endfloats*]{revtex4}

\usepackage{graphicx}
\usepackage{dcolumn}
\usepackage{bm}
\usepackage{amsfonts}

\DeclareMathAlphabet{\mathsfsl}{OT1}{cmr}{bx}{it}
\begin{document}
\title{Aging and rejuvenation during elastostatic loading of amorphous alloys}
\author{Nikolai V. Priezjev$^{1,2}$}
\affiliation{$^{1}$Department of Mechanical and Materials
Engineering, Wright State University, Dayton, OH 45435}
\affiliation{$^{2}$National Research University Higher School of
Economics, Moscow 101000, Russia}
\date{\today}
\begin{abstract}

Using molecular dynamics simulations, we investigate the effect of
uniaxial elastostatic compression on the potential energy,
structural relaxation, and mechanical properties of binary glasses.
We consider the three-dimensional Kob-Andersen binary mixture, which
was initially cooled from the liquid state to the glass phase with a
slow rate at zero pressure. The glass was then loaded with a static
stress at the annealing temperature during extended time intervals.
It is found that the static stress below the yielding point induces
large-scale plastic deformation and significant rejuvenation when
the annealing temperature is smaller than a fraction of the glass
transition temperature.  By contrast, aging effects become dominant
at sufficiently small values of the static stress and higher
annealing temperatures.  The mechanical tests after the elastostatic
compression have shown that both the elastic modulus and the yield
stress decrease in rejuvenated samples, while the opposite trend was
observed for relaxed glasses.  These results might be useful for the
thermomechanical processing of metallic glasses with optimized
mechanical properties.

\vskip 0.5in

Keywords: metallic glasses, deformation, mechanical treatment, yield
stress, molecular dynamics simulations

\end{abstract}

\maketitle

\section{Introduction}

The development and implementation of novel processing techniques
for metallic glasses in order to optimize their mechanical
properties is important for numerous structural and biomedical
applications~\cite{Greer16}.  It is well known however that in
contrast to crystalline materials, plastic deformation of metallic
glasses is associated with the formation of very localized shear
bands, thus preventing their widespread use~\cite{ Greer16}. In
practice, improved mechanical properties can be achieved by
relocating metallic glasses to higher potential energy states using
various thermal and mechanical treatments, such as cold rolling,
shot peening, high pressure torsion, ion irradiation, and, more
recently, cryogenic thermal cycling~\cite{Ketov15,Guo19}.
Alternatively, in cases when the shape change due to flow is not
important, higher energy states can be attained by reheating
amorphous alloys above the glass transition temperature followed by
suitably fast cooling back to the glass
state~\cite{Ogata15,Maass18,Yang18,Priez19one}.  Somewhat
unexpectedly, a convenient and efficient method to enhance the
potential energy of a glass is to simply apply a static stress in
the elastic range and let the system accumulate irreversible
structural
transformations~\cite{Park08,Jae-Chul09,Tong13,Wang15,Bai15,GreerSun16,Zhang17,PanGreer18,Samavatian19}.
Thus, it was demonstrated experimentally that the elastostatic
compression of Zr-based bulk metallic glasses leads to a significant
increase in enthalpy, which is comparable with the states produced
upon very fast quenching from the liquid state or by severe plastic
deformation~\cite{PanGreer18}.  Furthermore, upon increasing stress,
a transition from relaxation to rejuvenation in metallic glasses
subjected to elastostatic compression was observed~\cite{Zhang17}.
Despite recent progress, however, the mechanisms of rejuvenation as
well as the details of the elastostatic loading protocol required
for the maximum energy storage remain not fully understood.

\vskip 0.05in

In recent years, atomistic simulations were particularly useful in
elucidating the relaxation dynamics of amorphous allows subjected to
various deformation protocols including oscillatory shear
strain~\cite{IdoNature15,Priezjev16,Priezjev16a,Sastry17,Priezjev17,Priezjev18,Priezjev18a,Regev18,SastryPRX19},
constant~\cite{Yip17} and periodic~\cite{Lo10,NVP18strload} elastic
stress, as well as thermal cycling~\cite{Priez18tcyc,Priez19T2000}.
It was originally found that physical aging below the glass
transition temperature can be reversed by plastic
deformation~\cite{Stillinger00}, while the glass becomes effectively
overaged when reversibly deformed below yield~\cite{Lacks04}.
Interestingly, the structural relaxation slows down in case of
multiple subyield cycles, leading to progressively lower potential
energy states as the cycle number
increases~\cite{Sastry17,Priezjev18,Priezjev18a,NVP18strload}. On
the other hand, the yielding transition occurs after a certain
number of transient cycles if the strain amplitude is sufficiently
large, and the increase in the potential energy is caused by the
formation of a system-spanning shear
band~\cite{Sastry17,Priezjev17,Priezjev18a}.  However, the precise
control of the degree of relaxation or rejuvenation during
processing of metallic glasses remains a challenging task due to the
vast parameter space involved in loading protocols.

\vskip 0.05in

In this paper, molecular dynamics simulations are performed to
examine the effect of elastostatic loading on the potential energy
states and mechanical properties of an amorphous alloy.  After slow
initial cooling to a temperature well below the glass transition
temperature, the binary glass is loaded with a static stress in the
elastic range at the annealing temperature for extended periods of
time.   It will be shown that aging effects are predominant at
sufficiently low values of the applied stress when the annealing
temperature is greater than approximately half $T_g$.  By contrast,
high compressive stress (below yield) induces large-scale plastic
deformation and significant rejuvenation at low annealing
temperatures.   As a result of the treatment, aged/rejuvenated
samples exhibit higher/lower yield stress and elastic modulus during
strain-controlled compression.

\vskip 0.05in

This paper is structured as follows. The parameters of the molecular
dynamics simulation model and the details of the loading protocol
are given in the next section. The dependence of the potential
energy, elastic modulus, and yield stress on the annealing time,
temperature, and static stress, as well as spatial configurations of
mobile atoms are presented in section\,\ref{sec:Results}.  The
results are briefly summarized in the last section.

\section{Molecular dynamics simulations}
\label{sec:MD_Model}

The amorphous alloy is modeled using the Lennard-Jones (LJ) binary
(80:20) mixture first introduced over twenty years ago by Kob and
Andersen (KA)~\cite{KobAnd95}.   The LJ parameters of the KA model
is similar to the parametrization proposed by Weber and Stillinger
to simulate the amorphous metal alloy
$\text{Ni}_{80}\text{P}_{20}$~\cite{Weber85}. In the KA model, any
two atoms of types $\alpha,\beta=A,B$ interact via the truncated LJ
potential:
\begin{equation}
V_{\alpha\beta}(r)=4\,\varepsilon_{\alpha\beta}\,\Big[\Big(\frac{\sigma_{\alpha\beta}}{r}\Big)^{12}\!-
\Big(\frac{\sigma_{\alpha\beta}}{r}\Big)^{6}\,\Big],
\label{Eq:LJ_KA}
\end{equation}
with the parameters: $\varepsilon_{AA}=1.0$, $\varepsilon_{AB}=1.5$,
$\varepsilon_{BB}=0.5$, $\sigma_{AA}=1.0$, $\sigma_{AB}=0.8$,
$\sigma_{BB}=0.88$, and $m_{A}=m_{B}$~\cite{KobAnd95}.   This choice
of parameters corresponds to strongly non-additive interaction
between atoms $A$ and $B$, which prevents crystallization when the
binary mixture is cooled below the glass transition
temperature~\cite{KobAnd95}. To reduce the computational demand, the
cutoff radius of the LJ potential is set to
$r_{c,\,\alpha\beta}=2.5\,\sigma_{\alpha\beta}$ throughout the
study. The total number of atoms is $N=60\,000$.   The LJ units of
length, mass, energy, and time are adopted for all physical
quantities as follows: $\sigma=\sigma_{AA}$, $m=m_{A}$,
$\varepsilon=\varepsilon_{AA}$, and
$\tau=\sigma\sqrt{m/\varepsilon}$. All simulations were carried out
using the LAMMPS software~\cite{Lammps} with the integration time
step $\triangle t_{MD}=0.005\,\tau$.

\vskip 0.05in

\begin{figure}[t]
\includegraphics[width=9.0cm,angle=0]{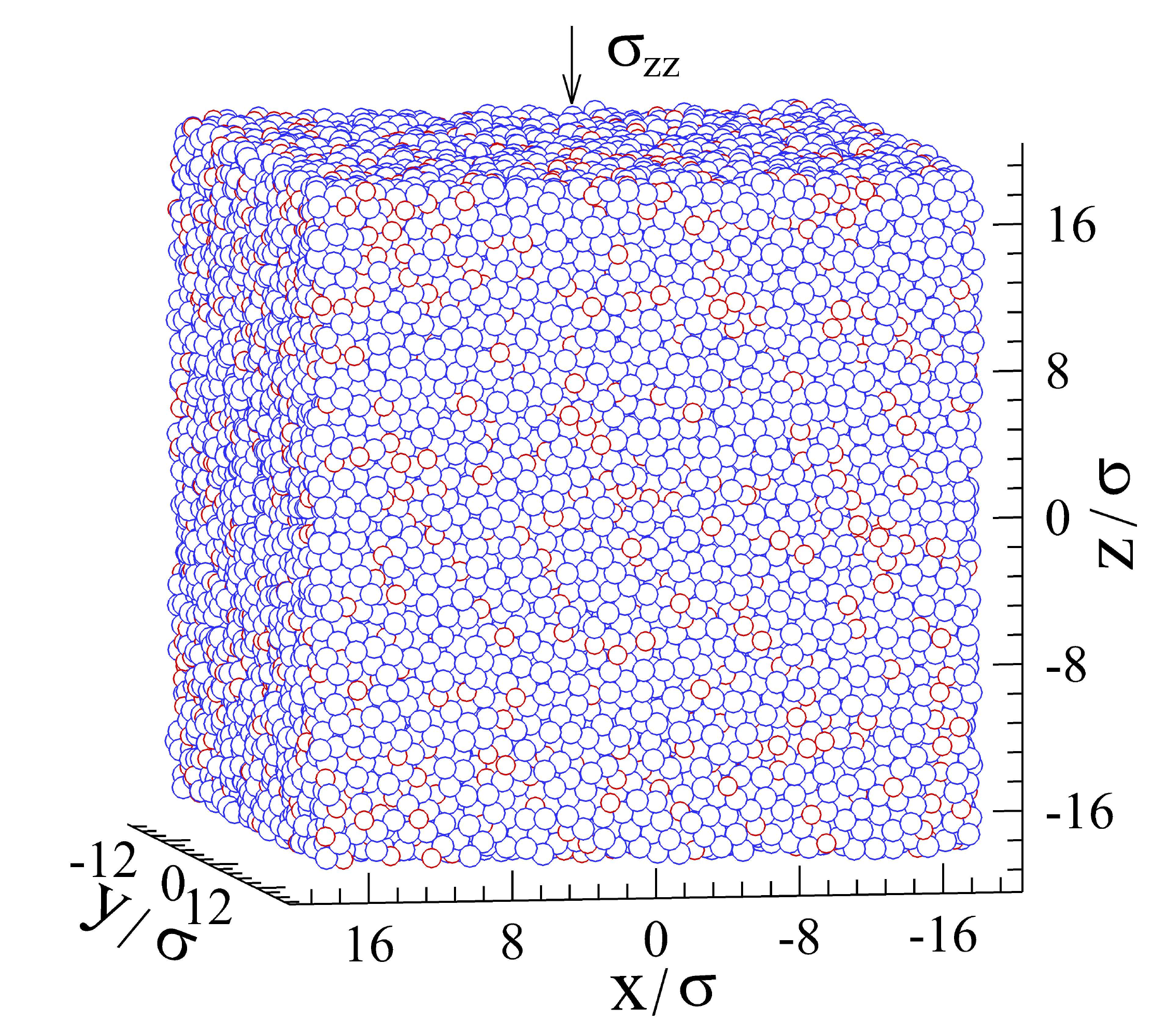}
\caption{(Color online) The spatial configuration of LJ atoms after
cooling from $T_{LJ}=1.0\,\varepsilon/k_B$ to
$0.01\,\varepsilon/k_B$ with the rate $10^{-5}\varepsilon/k_{B}\tau$
at zero pressure. The total number of atoms is $60\,000$. Note that
atoms of types $A$ and $B$ are indicated by large blue and small red
spheres, respectively. During elastostatic loading, the normal
stress $\sigma_{zz}$ is applied along the $\hat{z}$ direction, while
stresses $\sigma_{xx}$ and $\sigma_{yy}$ are set to zero.}
\label{fig:snapshot_system}
\end{figure}

The binary glass was first prepared by equilibrating the mixture at
the temperature $T_{LJ}=1.0\,\varepsilon/k_B$ and zero pressure,
using periodic boundary conditions. Here, $k_B$ and $T_{LJ}$ denote
the Boltzmann constant and temperature, respectively.  The
temperature was controlled via the Nos\'{e}-Hoover
thermostat~\cite{Allen87,Lammps}. Then, the binary mixture was
gradually cooled with the computationally slow rate
$10^{-5}\varepsilon/k_{B}\tau$ at zero pressure to the reference
temperature of $0.01\,\varepsilon/k_B$.  A snapshot of the amorphous
alloy at $T_{LJ}=0.01\,\varepsilon/k_B$ and $P=0$ is shown in
Fig.\,\ref{fig:snapshot_system}.

\vskip 0.05in


Next, the loading protocol was implemented via heating and at the
same time applying the normal stress along the $\hat{z}$ direction
during $5000\,\tau$. Then, the system was allowed to evolve during
the time period $t_{a}$ at the annealing temperature $T_{a}$ and the
static normal stress along the $\hat{z}$ direction, while the
pressure along the $\hat{x}$ and $\hat{y}$ directions were set to
zero. The last step is the decrease of temperature and normal stress
to the reference state, \textit{i.e.},
$T_{LJ}=0.01\,\varepsilon/k_B$ and $P=0$ in all three dimensions.
Thus, the control parameters of the elastostatic loading process
include the annealing temperature, the normal stress, and the
annealing time.  During each step of the process, the temperature,
stress components, potential energy, and system dimensions were
saved for further analysis. In addition, the mechanical properties
were probed by imposing the compressive strain along the $\hat{x}$
direction with the constant rate
$\dot{\varepsilon}_{xx}=10^{-5}\,\tau^{-1}$ at
$T_{LJ}=0.01\,\varepsilon/k_B$ and $P=0$. Due to computational
limitations, the simulations were performed for only one realization
of disorder.

\section{Results}
\label{sec:Results}


It is well recognized by now that the structural and mechanical
properties of metallic glasses depend sensitively on the rate of
cooling from the liquid state~\cite{Stillinger00}. Although in all
cases the glass structure is amorphous, more slowly annealed glasses
settle down into deeper energy minima and their global yield stress
becomes higher~\cite{Greer16}.  A more subtle characteristic of the
quenched glass is a broad distribution of the local yield stresses,
whose peak is displaced to higher values in more slowly annealed
glasses~\cite{Barbot18}.  Loosely speaking, a well annealed glass
can be viewed as a rigid matrix with randomly embedded soft spots
that have relatively low local yield stresses~\cite{GreerSun16}.
When a static stress below the global yield is applied, the matrix
is elastically deformed but the soft spots are prone to plastic
deformation over time, which might lead to rejuvenation and, as a
result, to improved plasticity. In what follows, we explore a wide
range of parameters, including the annealing temperature, time and
static stress, and observe both aging and rejuvenation in well
annealed binary glasses.

\vskip 0.05in

\begin{figure}[t]
\includegraphics[width=12.0cm,angle=0]{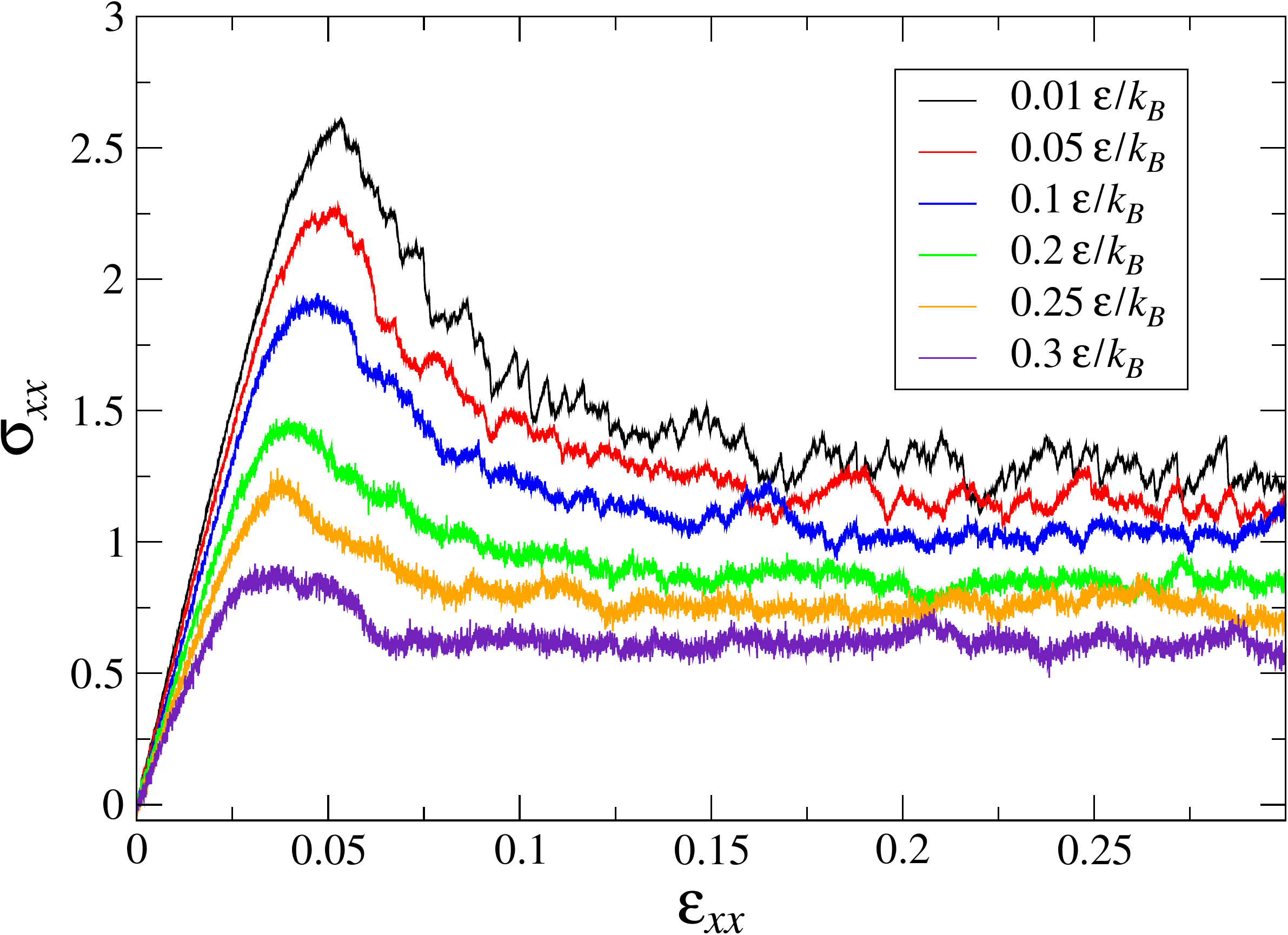}
\caption{(Color online) The dependence of stress (in units of
$\varepsilon\sigma^{-3}$) as a function of strain,
$\varepsilon_{xx}$, during compression along the $\hat{x}$ direction
with the rate $\dot{\varepsilon}_{xx}=10^{-5}\,\tau^{-1}$ at
temperatures $T_{LJ}=0.01\,\varepsilon/k_B$ (black),
$0.05\,\varepsilon/k_B$ (red), $0.1\,\varepsilon/k_B$ (blue),
$0.2\,\varepsilon/k_B$ (green), $0.25\,\varepsilon/k_B$ (orange),
and $0.3\,\varepsilon/k_B$ (indigo). The strain was imposed after
the sample was cooled below the glass transition temperature with
the rate $10^{-5}\varepsilon/k_{B}\tau$ at zero pressure. }
\label{fig:stress-strain_ini}
\end{figure}

It was previously shown that the glass transition temperature of the
KA binary mixture is $T_g\approx0.35\,\varepsilon/k_B$, which was
estimated by extrapolating the potential energy from the low and
high temperature regions when the system was cooled with the rate
$10^{-5}\varepsilon/k_{B}\tau$ at zero pressure~\cite{Priez19one}.
In order to determine the values of the stress overshot at different
temperatures, the compressive strain was applied to samples right
after cooling from the liquid state, \textit{i.e.}, to untreated
samples. The stress-strain curves are presented in
Fig.\,\ref{fig:stress-strain_ini} for selected values of $T_{LJ}$
lower than $T_g$.   As expected, upon decreasing $T_{LJ}$, the peak
value of the stress overshoot becomes higher and the corresponding
yield strain increases.  In the analysis below, the elastic modulus
$E=60.73\,\varepsilon\sigma^{-3}$ and the stress overshoot
$\sigma_Y=2.61\,\varepsilon\sigma^{-3}$, extracted from the
stress-strain curve at $T_{LJ}=0.01\,\varepsilon/k_B$, will serve as
reference values for comparison with treated samples.

\vskip 0.05in

\begin{figure}[t]
\includegraphics[width=12.0cm,angle=0]{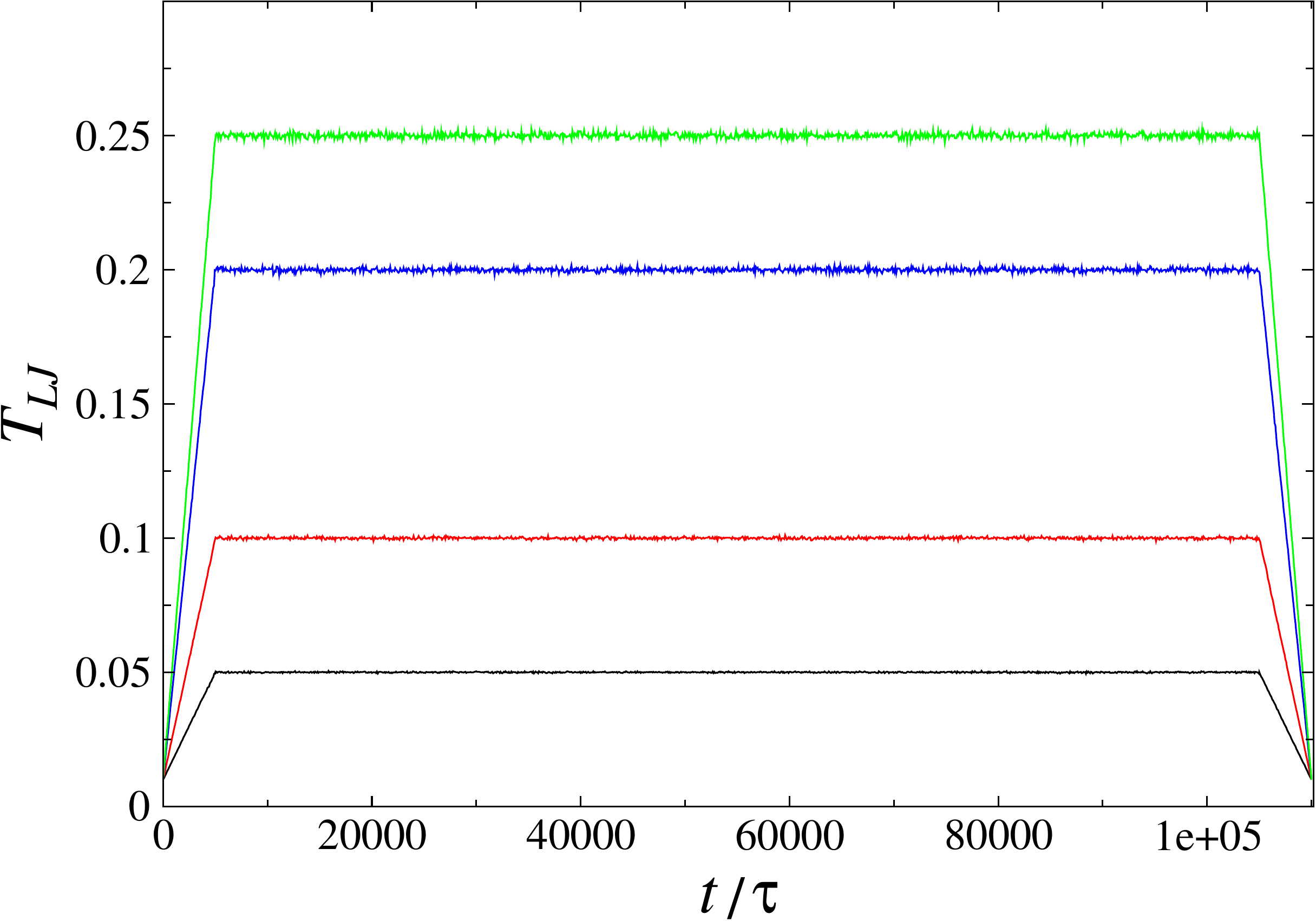}
\caption{(Color online) The temperature profiles during the loading
protocols, which include a gradual heating during $5000\,\tau$ from
$T_{LJ}=0.01\,\varepsilon/k_B$ to the annealing temperature,
$T_{a}$, followed by the annealing period $t_{a}=10^5\tau$, and a
subsequent cooling to the reference temperature
$T_{LJ}=0.01\,\varepsilon/k_B$ during $5000\,\tau$.  The values of
the annealing temperature are $T_{a}=0.05\,\varepsilon/k_B$ (black),
$0.1\,\varepsilon/k_B$ (red), $0.2\,\varepsilon/k_B$ (blue), and
$0.25\,\varepsilon/k_B$ (green).}
\label{fig:temper_profs}
\end{figure}

We next describe the elastostatic loading protocol and discuss the
choice of parameter values for the static stress, annealing
temperature, and the reference state. After the annealing time
interval, the glass was always brought to the reference state at
$T_{LJ}=0.01\,\varepsilon/k_B$ and $P=0$ to facilitate comparison of
the potential energy and mechanical properties. For example, the
variation of temperature during the loading process is illustrated
in Fig.\,\ref{fig:temper_profs} for the annealing time
$t_{a}=10^5\tau$. It can be seen that during the first $5000\,\tau$,
the temperature was ramped linearly to $T_a$ and kept constant
during the annealing time $t_a$. At the same time, the normal
stress, $\sigma_{zz}$, increased from zero to a finite value during
the first $5000\,\tau$ (not shown) and remained constant during
$t_a$, while the other stress components
$\sigma_{xx}=\sigma_{yy}=0$. After the time interval $t_a$, both the
temperature and normal stress were reduced back to their values at
the reference state ($T_{LJ}=0.01\,\varepsilon/k_B$ and
$\sigma_{zz}=0$) during $5000\,\tau$.   Once at reference state, the
potential energy was computed, and the sample was compressed along
the $\hat{x}$ direction with a constant strain rate to measure the
stress response (to be discussed below).

\vskip 0.05in

\begin{figure}[t]
\includegraphics[width=12.0cm,angle=0]{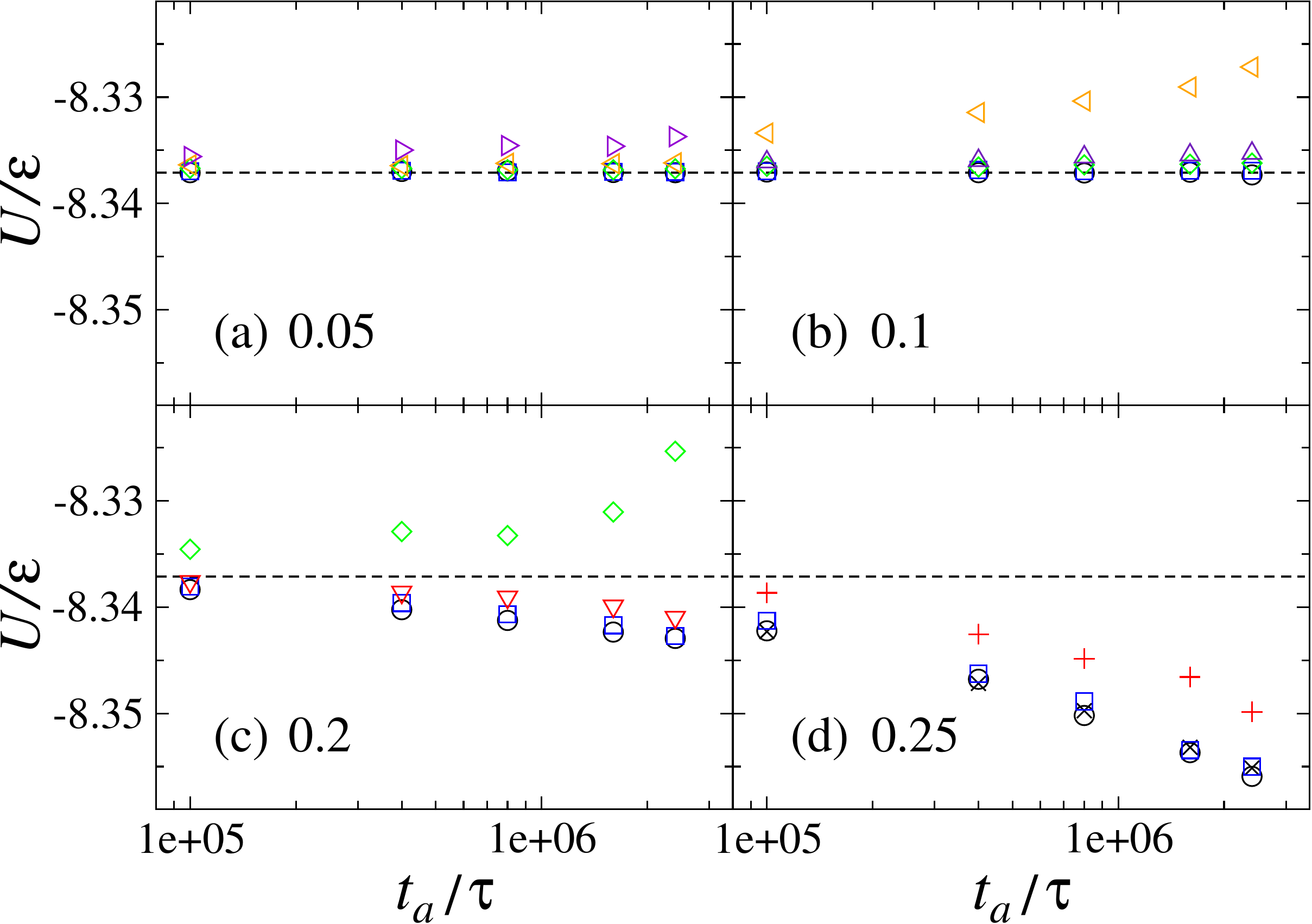}
\caption{(Color online) The potential energy as a function of the
annealing time, $t_{a}/\tau$, for static stresses $\sigma_{zz}=0.0$
($\circ$), $0.3\,\varepsilon\sigma^{-3}$ ($\times$),
$0.5\,\varepsilon\sigma^{-3}$ ($\square$),
$0.7\,\varepsilon\sigma^{-3}$ (+), $0.8\,\varepsilon\sigma^{-3}$
($\triangledown$), $1.0\,\varepsilon\sigma^{-3}$ ($\lozenge$),
$1.2\,\varepsilon\sigma^{-3}$ ($\vartriangle$),
$1.5\,\varepsilon\sigma^{-3}$ ($\triangleleft$), and
$1.8\,\varepsilon\sigma^{-3}$ ($\triangleright$). The annealing
temperatures are (a) $T_{a}=0.05\,\varepsilon/k_B$, (b)
$0.1\,\varepsilon/k_B$, (c) $0.2\,\varepsilon/k_B$, and (d)
$0.25\,\varepsilon/k_B$.  The horizontal dashed lines denote the
potential energy level $U=-8.337\,\varepsilon$ at
$T_{LJ}=0.01\,\varepsilon/k_B$ and $P=0$ before the thermomechanical
processing. }
\label{fig:poten_4_Ta}
\end{figure}

The dependence of the potential energy as a function of the
annealing time is shown in Fig.\,\ref{fig:poten_4_Ta} for the
annealing temperatures $T_{a}=0.05\,\varepsilon/k_B$,
$0.1\,\varepsilon/k_B$, $0.2\,\varepsilon/k_B$, and
$0.25\,\varepsilon/k_B$.   For each $T_{a}$, the simulations were
performed for several values of the static stress, $\sigma_{zz}$,
smaller than the stress overshoot reported in
Fig.\,\ref{fig:stress-strain_ini}. Note that the potential energy
level before the treatment, $U=-8.337\,\varepsilon$, is denoted by
the dashed lines in Fig.\,\ref{fig:poten_4_Ta}.   It is readily
apparent that the aging effects at $\sigma_{zz}=0$ are negligible
for $T_{a}\leqslant0.1\,\varepsilon/k_B$, whereas $U$ decreases
monotonically at higher temperatures $T_{a}=0.2\,\varepsilon/k_B$
and $0.25\,\varepsilon/k_B$ during $t_a=2.4\times10^6\tau$.

\vskip 0.05in


The influence of the static stress is reflected in the increase of
the potential energy with respect to $U(T_a, t_a)$ at
$\sigma_{zz}=0$, as shown in Fig.\,\ref{fig:poten_4_Ta}.   The most
pronounced increase in $U$ occurs when the static stress is
sufficiently close to the yield stress. For example, the most
rejuvenated states at $T_{a}=0.1\,\varepsilon/k_B$ are obtained when
$\sigma_{zz}=1.5\,\varepsilon\sigma^{-3}$ in
Fig.\,\ref{fig:poten_4_Ta}\,(b), while the peak value of the stress
overshoot at $T_{LJ}=0.1\,\varepsilon/k_B$ in
Fig.\,\ref{fig:stress-strain_ini} is about
$1.9\,\varepsilon\sigma^{-3}$.  In general, it is difficult to
identify a threshold value of the static stress that leads to
maximum rejuvenation and at the same time does not cause significant
deformation of the simulation domain over time.  For instance, when
the glass was loaded with $\sigma_{zz}=1.2\,\varepsilon\sigma^{-3}$
at $T_{a}=0.2\,\varepsilon/k_B$, the system became fully compressed
along the $\hat{z}$ direction after about $3.5\times10^4\tau$,
despite that the magnitude of the yielding peak is
$\approx\!1.43\,\varepsilon\sigma^{-3}$ at
$T_{LJ}=0.2\,\varepsilon/k_B$ in Fig.\,\ref{fig:stress-strain_ini}.
Furthermore, the transition from aging to rejuvenation upon
increasing $\sigma_{zz}$ is seen most clearly at
$T_{a}=0.2\,\varepsilon/k_B$ in Fig.\,\ref{fig:poten_4_Ta}\,(c).  We
finally comment that static loading at $T_{a}=0.25\,\varepsilon/k_B$
only reduces the effect of aging but the potential energy remains
below the level for the untreated sample, as shown in
Fig.\,\ref{fig:poten_4_Ta}\,(d).    Similarly, test runs at the
higher annealing temperature $T_{a}=0.3\,\varepsilon/k_B$ indicated
that $U$ also decreases over time for any values of the applied
static stress (not shown).

\vskip 0.05in


The details of the structural relaxation process can be unveiled by
analyzing the so-called nonaffine displacements of
atoms~\cite{Falk98}. In other words, the temporal evolution of the
system can be visualized as a sequence of the relative displacements
of atoms with respect to their neighbors. More specifically, the
nonaffine measure for the $i$-th atom is defined by computing the
transformation matrix $\mathbf{J}_i$, which performs a linear
transformation of its neighboring atoms during the time interval
$\Delta t$ and minimizes the following expression:
\begin{equation}
D^2(t, \Delta t)=\frac{1}{N_i}\sum_{j=1}^{N_i}\Big\{
\mathbf{r}_{j}(t+\Delta t)-\mathbf{r}_{i}(t+\Delta t)-\mathbf{J}_i
\big[ \mathbf{r}_{j}(t) - \mathbf{r}_{i}(t)    \big] \Big\}^2,
\label{Eq:D2min}
\end{equation}
where the sum is taken over the atoms within a sphere of radius
$1.5\,\sigma$ centered at $\mathbf{r}_{i}(t)$.  In particular, it
was recently shown that the amplitude of nonaffine displacements
during oscillatory deformation of amorphous solids is approximately
power-law distributed, and most of the atoms undergo reversible
displacements after one or several cycles when the strain amplitude
is below the yielding point~\cite{Priezjev16,Priezjev16a}.
Furthermore, the formation of shear bands at the yielding transition
was readily detected by identifying atoms with relatively large
nonaffine displacements (larger than the typical cage size) during
one shear cycle in both poorly~\cite{Priezjev18a} and
well~\cite{Priezjev17} annealed glasses.   In general, the nonaffine
measure is also an excellent diagnostic of local structural
rearrangements during
mechanical~\cite{Priezjev18,Priezjev18a,NVP18strload} and
thermal~\cite{Priez18tcyc} annealing processes, when disordered
solids gradually evolve towards low potential energy states.

\vskip 0.05in

\begin{figure}[t]
\includegraphics[width=12.cm,angle=0]{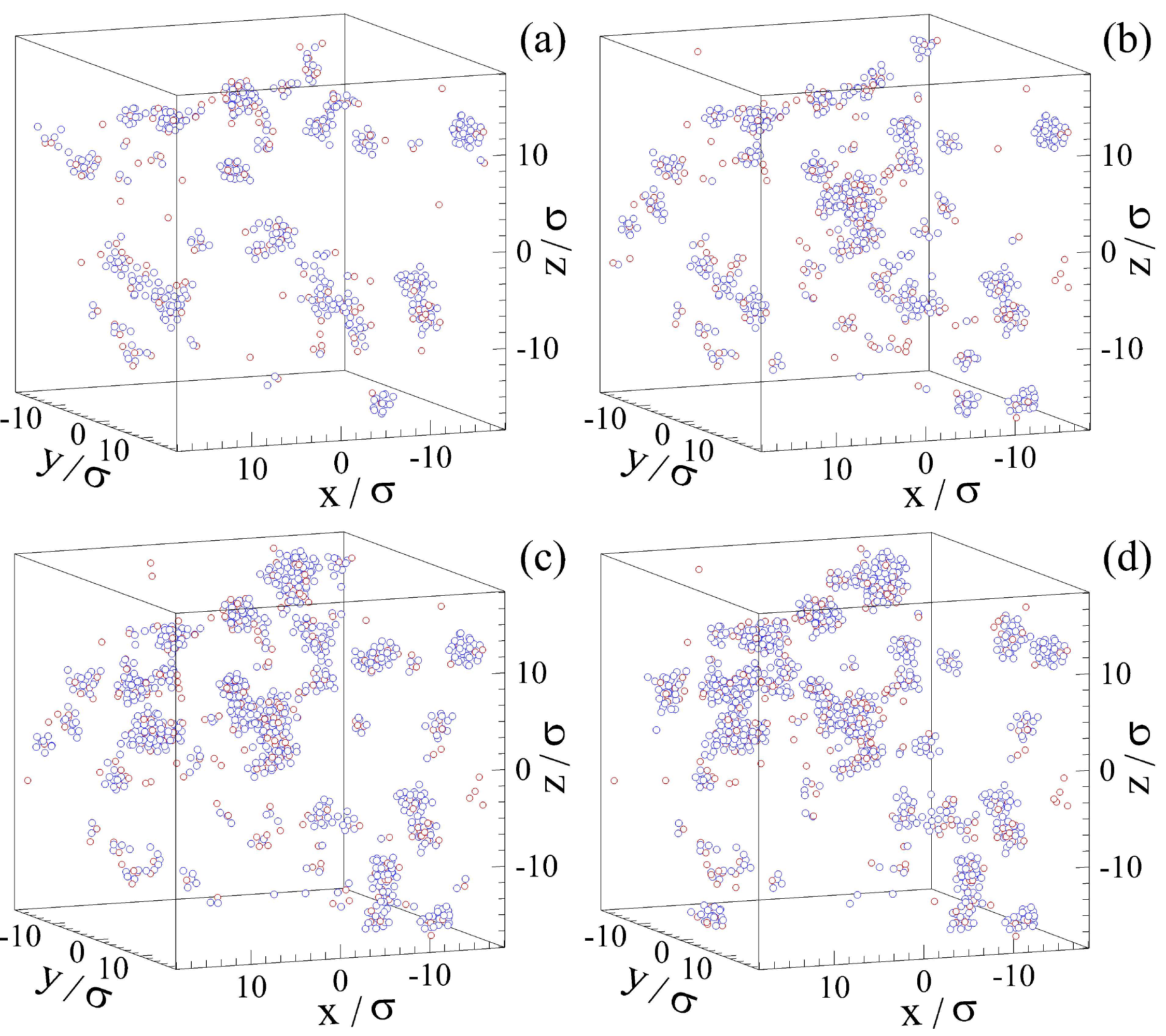}
\caption{(Color online) The snapshots of atoms with the nonaffine
measure: (a) $D^2(0, 10^5\tau)>0.04\,\sigma^2$, (b) $D^2(0,
4\times10^5\tau)>0.04\,\sigma^2$, (c) $D^2(0,
1.6\times10^6\tau)>0.04\,\sigma^2$, and (d) $D^2(0,
2.4\times10^6\tau)>0.04\,\sigma^2$. The sample was kept at the
temperature $T_{a}=0.1\,\varepsilon/k_B$ and the static stress
$\sigma_{zz}=0.5\,\varepsilon\sigma^{-3}$. }
\label{fig:snapshot_clusters_T01_Pz05}
\end{figure}
\begin{figure}[t]
\includegraphics[width=12.cm,angle=0]{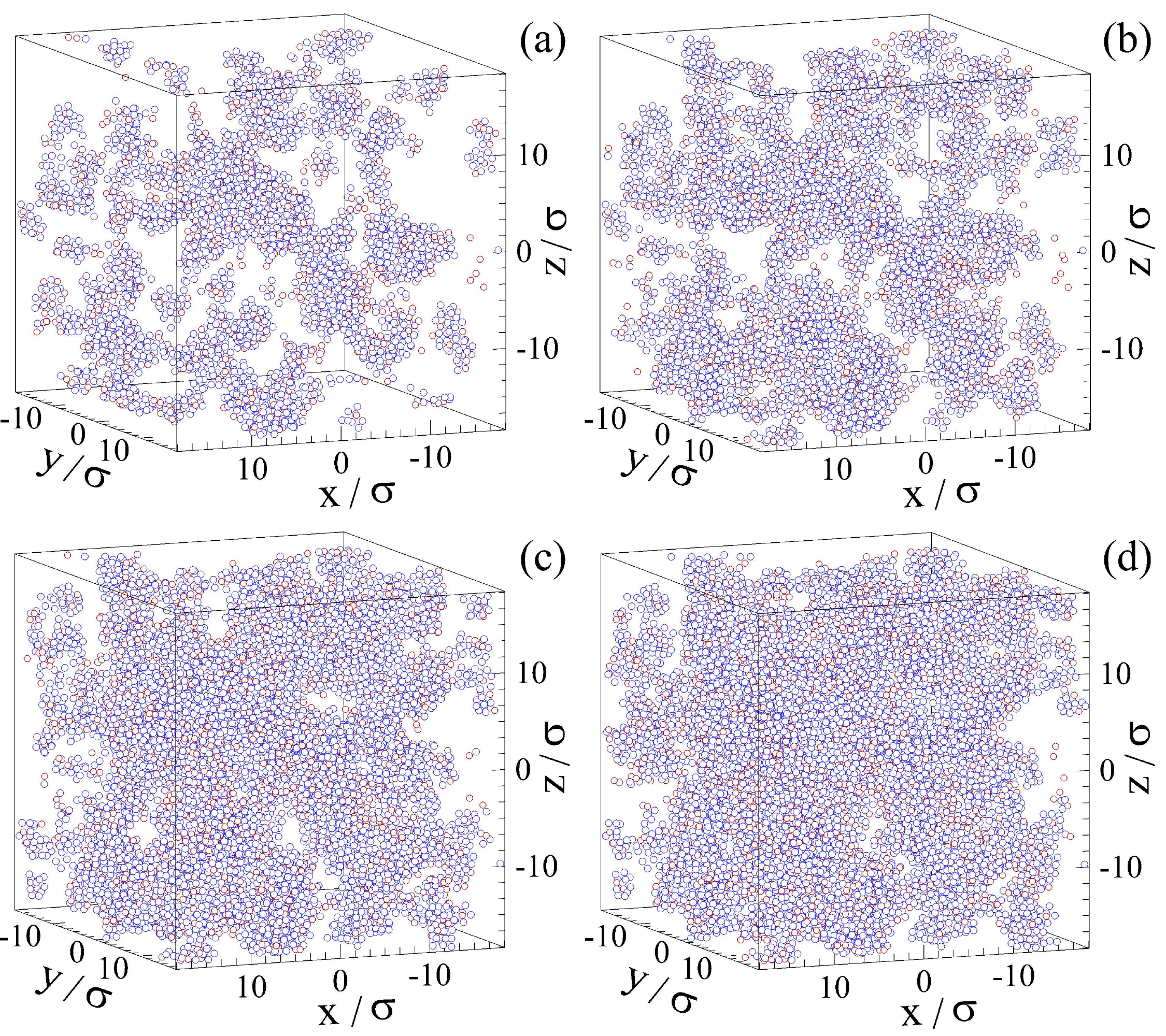}
\caption{(Color online) The positions of atoms with large nonaffine
measure: (a) $D^2(0, 10^5\tau)>0.04\,\sigma^2$, (b) $D^2(0,
4\times10^5\tau)>0.04\,\sigma^2$, (c) $D^2(0,
1.6\times10^6\tau)>0.04\,\sigma^2$, and (d) $D^2(0,
2.4\times10^6\tau)>0.04\,\sigma^2$. The annealing temperature is
$T_{a}=0.1\,\varepsilon/k_B$ and the static stress is
$\sigma_{zz}=1.5\,\varepsilon\sigma^{-3}$. }
\label{fig:snapshot_clusters_T01_Pz15}
\end{figure}

The representative spatial configurations of atoms with relatively
large nonaffine measure, $D^2(0, t_{a})>0.04\,\sigma^2$, are
displayed in Fig.\,\ref{fig:snapshot_clusters_T01_Pz05} for
$\sigma_{zz}=0.5\,\varepsilon\sigma^{-3}$ and in
Fig.\,\ref{fig:snapshot_clusters_T01_Pz15}  for
$\sigma_{zz}=1.5\,\varepsilon\sigma^{-3}$ during loading at
$T_{a}=0.1\,\varepsilon/k_B$.   To remind, the analysis of nonaffine
displacements between two atomic configurations separated by the
time interval $t_{a}$ was performed when the system was brought to
the reference state at $T_{LJ}=0.01\,\varepsilon/k_B$ and $P=0$.
Since the typical cage size is $r_c\approx 0.1\,\sigma$,
Figs.\,\ref{fig:snapshot_clusters_T01_Pz05} and
\ref{fig:snapshot_clusters_T01_Pz15} only show atoms that left their
cages during $t_{a}$, and, therefore, they represent the local
plastic deformation of the material.    It can be observed in
Fig.\,\ref{fig:snapshot_clusters_T01_Pz05} that mobile atoms are
organized into small clusters whose size slightly increases over
time, while the net effect of these transformations on the potential
energy is negligible [see the data for
$\sigma_{zz}=0.5\,\varepsilon\sigma^{-3}$ in
Fig.\,\ref{fig:poten_4_Ta}\,(b)]. By contrast, loading at
$T_{a}=0.1\,\varepsilon/k_B$ and
$\sigma_{zz}=1.5\,\varepsilon\sigma^{-3}$ induces extended clusters
that become comparable with the system size when $t_{a}\gtrsim
10^6\tau$ (see Fig.\,\ref{fig:snapshot_clusters_T01_Pz15}), which
leads to significant rejuvenation, as shown in
Fig.\,\ref{fig:poten_4_Ta}\,(b).  We comment that despite
large-scale structural rearrangements at
$\sigma_{zz}=1.5\,\varepsilon\sigma^{-3}$, the net compression along
the $\hat{z}$ direction is relatively small; \textit{i.e.}, from
$L_z=36.54\,\sigma$ at $t_{a}=0$ to $L_z=36.11\,\sigma$ at
$t_a=2.4\times10^6\tau$.

\vskip 0.05in


It should be emphasized that there is a crucial difference between
rejuvenated states obtained via elastostatic loading and cyclic
shear with the strain amplitude above the yielding point. In the
latter case, the increase in the potential energy is directly
related to the formation of a fluidized shear
band~\cite{Priezjev17,Sastry17,Priezjev18a,SastryPRX19}, while in
the former case, the plastic deformation under static stress is
distributed more homogeneously in the regions with relatively low
local yield stress.   The next comment is regarding the structural
characteristics at different energy states.  It was previously shown
that unlike Zr-based metallic glasses, where the icosahedral
short-range order is a sensitive indicator of structural changes,
the KA binary glass does not contain frequent icosahedral
structures, but instead its microscopic structure is rather
sensitive to the shape of the pair correlation function of small
atoms of type $B$~\cite{Vollmayr96,Stillinger00,Priez19one}. In the
recent study on thermal processing of binary glasses, the changes in
the potential energy were correlated with the height of the
nearest-neighbor peak in the $BB$ pair correlation
function~\cite{Priez19one}.  Since the variation of the potential
energy reported in Fig.\,\ref{fig:poten_4_Ta} is comparable with
energy changes found in Ref.\,\cite{Priez19one} and the variation of
the peak height was relatively small, the structural analysis was
not performed in the present study.

\vskip 0.05in

\begin{figure}[t]
\includegraphics[width=12.cm,angle=0]{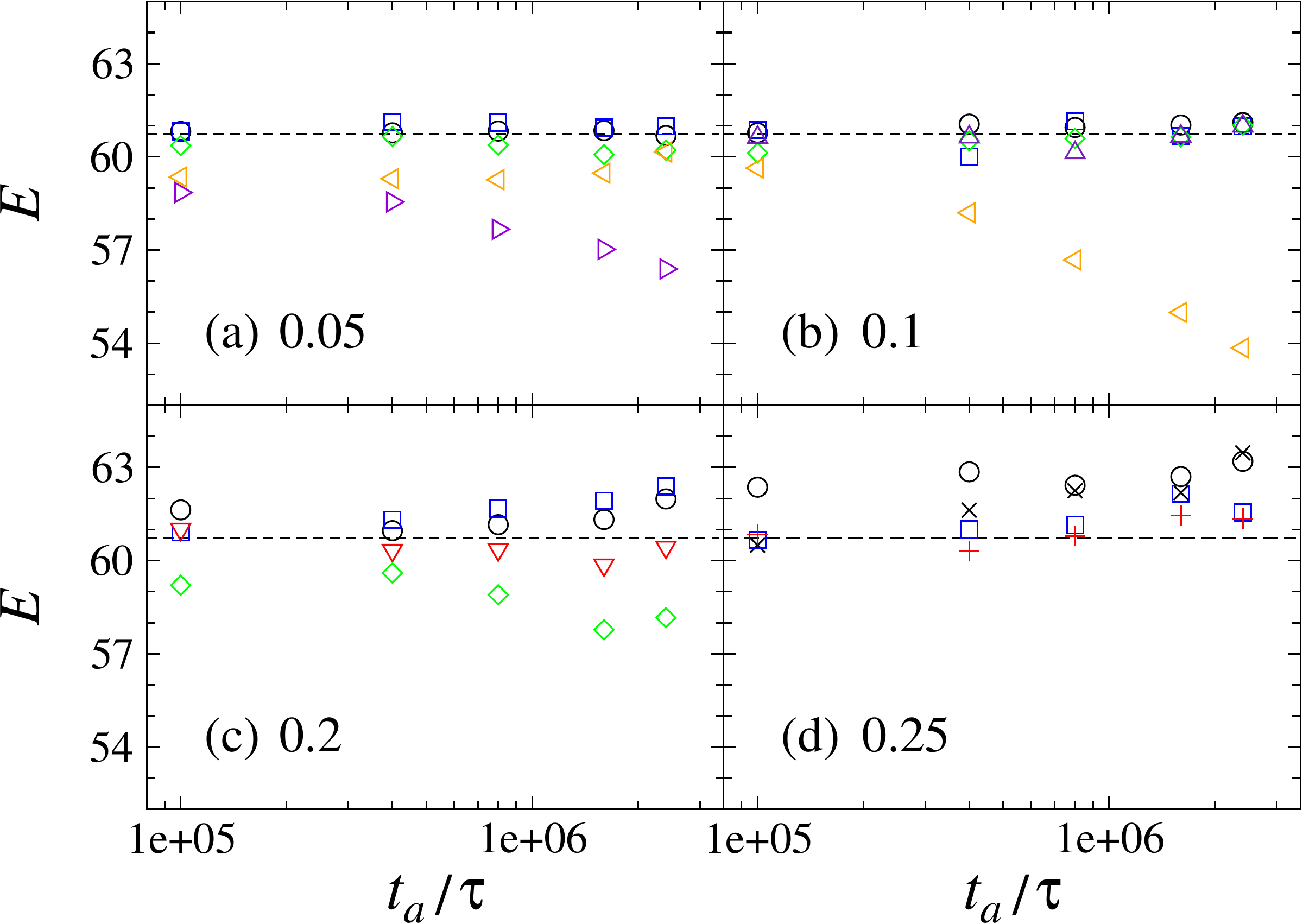}
\caption{(Color online) The elastic modulus $E$ (in units of
$\varepsilon\sigma^{-3}$) versus the annealing time, $t_{a}$, and
temperatures (a) $T_{a}=0.05\,\varepsilon/k_B$, (b)
$0.1\,\varepsilon/k_B$, (c) $0.2\,\varepsilon/k_B$, and (d)
$0.25\,\varepsilon/k_B$.  The black dashed lines indicate the value
of the elastic modulus $E=60.73\,\varepsilon\sigma^{-3}$ before the
elastostatic loading. The static stress is $\sigma_{zz}=0.0$
($\circ$), $0.3\,\varepsilon\sigma^{-3}$ ($\times$),
$0.5\,\varepsilon\sigma^{-3}$ ($\square$),
$0.7\,\varepsilon\sigma^{-3}$ (+), $0.8\,\varepsilon\sigma^{-3}$
($\triangledown$), $1.0\,\varepsilon\sigma^{-3}$ ($\lozenge$),
$1.2\,\varepsilon\sigma^{-3}$ ($\vartriangle$),
$1.5\,\varepsilon\sigma^{-3}$ ($\triangleleft$), and
$1.8\,\varepsilon\sigma^{-3}$ ($\triangleright$).}
\label{fig:E}
\end{figure}
\begin{figure}[t]
\includegraphics[width=12.cm,angle=0]{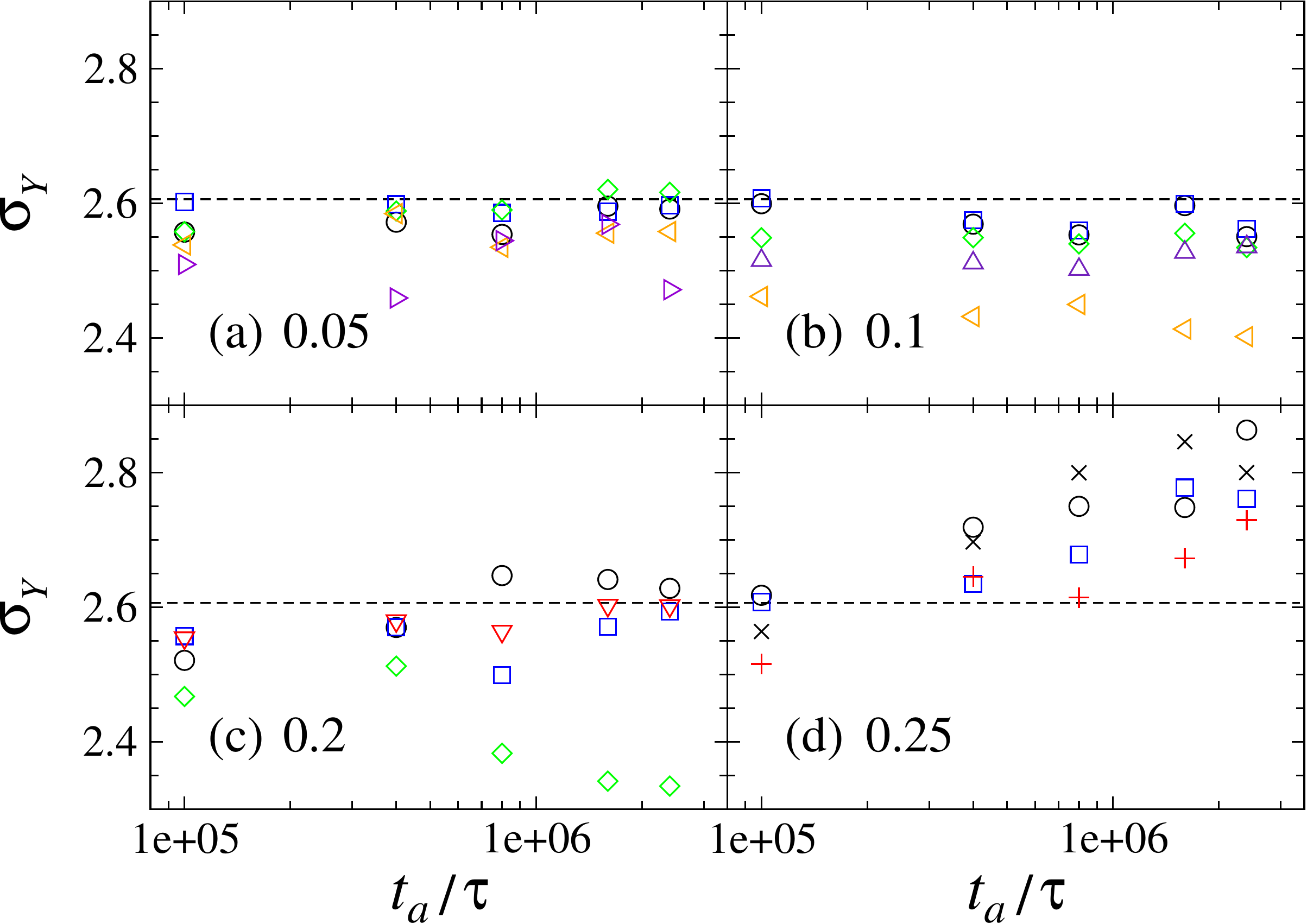}
\caption{(Color online) The yielding peak $\sigma_Y$ (in units of
$\varepsilon\sigma^{-3}$) as a function of the annealing time and
temperatures (a) $T_{a}=0.05\,\varepsilon/k_B$, (b)
$0.1\,\varepsilon/k_B$, (c) $0.2\,\varepsilon/k_B$, and (d)
$0.25\,\varepsilon/k_B$. The horizontal lines mark the value
$\sigma_Y=2.61\,\varepsilon\sigma^{-3}$ before the thermal loading.
The applied stress is $\sigma_{zz}=0.0$ ($\circ$),
$0.3\,\varepsilon\sigma^{-3}$ ($\times$),
$0.5\,\varepsilon\sigma^{-3}$ ($\square$),
$0.7\,\varepsilon\sigma^{-3}$ (+), $0.8\,\varepsilon\sigma^{-3}$
($\triangledown$), $1.0\,\varepsilon\sigma^{-3}$ ($\lozenge$),
$1.2\,\varepsilon\sigma^{-3}$ ($\vartriangle$),
$1.5\,\varepsilon\sigma^{-3}$ ($\triangleleft$), and
$1.8\,\varepsilon\sigma^{-3}$ ($\triangleright$). }
\label{fig:sigY}
\end{figure}

We finally report the elastic modulus in Fig.\,\ref{fig:E} and the
peak value of the stress overshoot in Fig.\,\ref{fig:sigY} as
functions of the annealing time and temperatures
$T_{a}=0.05\,\varepsilon/k_B$, $0.1\,\varepsilon/k_B$,
$0.2\,\varepsilon/k_B$, and $0.25\,\varepsilon/k_B$.  In each case,
the stress was computed during compression along the $\hat{x}$
direction with the rate $\dot{\varepsilon}_{xx}=10^{-5}\,\tau^{-1}$
at $T_{LJ}=0.01\,\varepsilon/k_B$ and $\sigma_{zz}=\sigma_{yy}=0$.
Although the data are somewhat scattered, it is evident in
Figs.\,\ref{fig:E} and \ref{fig:sigY} that aged/rejuvenated glasses
exhibit higher/lower values of $E$ and $\sigma_Y$. The most
pronounced decrease in the elastic modulus and yield stress occurs
for the following parameters of the loading protocol ($T_a$,
$\sigma_{zz}$): ($0.05\,\varepsilon/k_B$,
$1.8\,\varepsilon\sigma^{-3}$), ($0.1\,\varepsilon/k_B$,
$1.5\,\varepsilon\sigma^{-3}$), and ($0.2\,\varepsilon/k_B$,
$1.0\,\varepsilon\sigma^{-3}$). Interestingly, a similar decrease in
$E$ and $\sigma_Y$ (about $10\%$) was reported in the recent study
where a glass was reheated above $1.3\,T_g$ and then quenched with a
rate of about $10^{-3}\varepsilon/k_{B}\tau$ to the glass
phase~\cite{Priez19one}. We lastly mention that the increase in
$\sigma_Y$ as a function of $t_a$ at $\sigma_{zz}=0$ shown in
Fig.\,\ref{fig:sigY}\,(c,\,d) is consistent with the results of
previous studies on the effect of physical aging on the yield
stress~\cite{Varnik04,Rottler05}.


\section{Conclusions}

In summary, the influence of prolonged elastostatic compression on
the potential energy, structural relaxation, and mechanical
properties of binary glasses was examined using molecular dynamics
simulations. The model glass was represented via the Lennard-Jones
binary mixture, which was initially annealed from the liquid state
to the glass phase with a computationally slow rate at zero
pressure.  During the thermomechanical treatment, the glass was
loaded with a compressive stress and annealed at a constant
temperature.    It was found that the static stress below the yield
stress induces significant rejuvenation due to large-scale plastic
deformation.     In contrast, at sufficiently low stress levels,
aging becomes dominant when the annealing temperature is greater
than about half $T_g$.   After the elastostatic loading, both the
elastic modulus and the yield stress were found to decrease in
rejuvenated glasses, while the opposite trend was reported for aged
samples.

\section*{Acknowledgments}

Financial support from the National Science Foundation (CNS-1531923)
is gratefully acknowledged. The article was prepared within the
framework of the HSE University Basic Research Program and funded in
part by the Russian Academic Excellence Project `5-100'. The
molecular dynamics simulations were performed using the LAMMPS
software developed at Sandia National Laboratories~\cite{Lammps}.
The numerical simulations were performed at Wright State
University's Computing Facility and the Ohio Supercomputer Center.


%

%
%

%
%

%
%

%

%

%

%

\bibliographystyle{prsty}

\end{document}